\documentclass[conference]{IEEEtran}
\usepackage[utf8]{inputenc}

\usepackage[english]{babel}
\usepackage{amsmath}
\usepackage{amssymb}
\usepackage{amsfonts}
\usepackage{graphicx}
\usepackage{booktabs}      % for \toprule in tables
\usepackage{multirow}      % for table multirow
\usepackage{adjustbox}
\usepackage{threeparttable}
\usepackage{parcolumns}
\usepackage{tabularx}
\usepackage{diagbox}
\usepackage{xcolor}
\usepackage[absolute]{textpos}

\usepackage{newtxtext}   % replacement for times package
\usepackage{inconsolata} % changes the default typewriter font
\usepackage{microtype}
\DisableLigatures[f]{encoding = *, family = * }

\usepackage{hyperref}
\hypersetup{pdfborder={0 0 0},colorlinks=true,urlcolor=blue,linkcolor=blue,citecolor=blue,bookmarks=true}
\hypersetup{pdftitle={Resolving Ethics Trade-offs in Implementing Responsible AI}}
\hypersetup{pdfauthor={Conrad Sanderson, Emma Schleiger, David Douglas, Petra Kuhnert, Qinghua Lu}}

\usepackage{caption}
\usepackage[shortlabels]{enumitem}
\usepackage{balance}

\usepackage{wasysym}  % for \CIRCLE, \LEFTcircle

\usepackage[hang,flushmargin]{footmisc}   % remove indent in footnotes

\makeatletter
\def\footnoterule{\kern-3\p@
  \hrule \@width \columnwidth \kern 2.6\p@} % the \hrule is .4pt high
\makeatother

\newcolumntype{R}[2]{%
  >{\adjustbox{angle=#1,lap=\width-(#2)}\bgroup}%
  l%
  <{\egroup}%
}
\newcommand*\rot{\multicolumn{1}{R{35}{2.0em}}}

\def\andor{{\mbox{and{\hspace{0.1ex}}/or}}}
\def\AIML{\mbox{AI{\hspace{0.1ex}}/{\hspace{0.1ex}}ML}}

\begin{document}

\title{\scalebox{1.55}{\normalsize\textbf{Resolving Ethics Trade-offs in Implementing Responsible AI}}}
\author
  {
  Conrad Sanderson\textsuperscript{{\tiny~}$\dagger\diamond$},
  Emma Schleiger\textsuperscript{{\tiny~}$\dagger$},
  David Douglas\textsuperscript{{\tiny~}$\ddagger$},
  Petra Kuhnert\textsuperscript{{\tiny~}$\dagger$},
  Qinghua Lu\textsuperscript{{\tiny~}$\dagger$}\\
  ~\\
  \textsuperscript{$\dagger$}{\tiny~}\textit{Data61 / CSIRO, Australia;}~
  \textsuperscript{$\ddagger$}{\tiny~}\textit{Environment / CSIRO, Australia;}~
  \textsuperscript{$\diamond$}{\tiny~}\textit{Griffith University, Australia}
  }

\maketitle

\begin{abstract}

While the operationalisation of high-level AI ethics principles into practical {\AIML} systems has made progress,
there is still a theory-practice gap in managing tensions between the underlying AI ethics aspects.
We cover five approaches for addressing the tensions via trade-offs,
ranging from rudimentary to complex.
The approaches differ in the types of considered context,
scope, methods for measuring contexts,
and degree of justification.
None of the approaches is likely to be appropriate for all organisations, systems, or applications.
To address this, we propose a framework which consists of:
(i)~proactive identification of tensions,
(ii)~prioritisation and weighting of ethics aspects,
(iii)~justification and documentation of trade-off decisions.
The proposed framework aims to facilitate the implementation of well-rounded {\AIML} systems
that are appropriate for potential regulatory requirements.
\end{abstract}

\vspace{1ex}

\begin{IEEEkeywords}
AI ethics, responsible AI, tensions, trade-offs, system design, societal impact, governance, regulations.
\end{IEEEkeywords}

\begin{textblock}{13.44}(1.28,14.75)
\hrule
\vspace{1ex}
\noindent
\footnotesize
\textbf{{$^\ast$}~Published in:} IEEE Conference on Artificial Intelligence, 2024. DOI:~\href{https://doi.org/10.1109/CAI59869.2024.00215}{\tt 10.1109/CAI59869.2024.00215}
\end{textblock}

\section{Introduction}

The increasing impact of artificial intelligence~(AI) and machine learning~(ML) across
many sectors of society has led to a broad consensus on the need to design and 
implement these technologies in a responsible manner. 
Globally, governments, organisations and industry groups
have defined many sets of high-level AI ethics principles to support this vision~\cite{jobin2019global};
an~example is shown in Table~\ref{tab:ai_ethics_principles}.
Such high-level principles contain a set of underlying themes and aspects, 
which typically includes
accuracy{\hspace{0.1ex}}/{\hspace{0.1ex}}performance,
robustness{\hspace{0.2ex}}/{\hspace{0.1ex}}safety,
fairness, privacy,
explainability{\hspace{0.1ex}}/{\hspace{0.2ex}}interpretability,
transparency, and accountability \cite{Fjeld_2020,jobin2019global}.

As the high-level principles are implemented in practice, the underlying aspects interact,
which inevitably leads to various tensions and trade-offs~\cite{Bleher_2023,Sanderson_2023b,Whittlestone_2019}.
Among the many observed interactions (summarised in Table~\ref{tab:ethics_interactions}),
a prominent example is the trade-off between accuracy and explainability {\cite{Arrieta_2020,Holm_2019,Petkovic_2023}}.
While the need to formally address the \mbox{trade-offs} is often stated in the literature 
\cite{Huang_2023,Peters_2020,Smit_2020,Whittlestone_2019},
there is currently no generally agreed upon framework~to~accomplish~this.

A further complication is that many designers and developers of {\AIML} systems%
\footnote{In this work we use the term \textit{{\AIML} systems}
to refer to both ML models (algorithms) and AI products.}
are currently unaware of the tensions and trade-offs,
which may stem from unfamiliarity of
(or their unwillingness to engage with)
AI ethics principles
{\andor} their underlying aspects \cite{Vakkuri_2020,Varanasi_2023}.
Without regulatory enforcement,
taking AI ethics principles into account can be contrary to industry priorities~\cite{McLennan_2020}.
For example, it has been observed that taking into account the fairness aspect
can considerably reduce the accuracy of {\AIML} systems, affecting the potential
profitability to be gained from using these systems~\cite{Kozodoi_2022}.

A~recent analysis of highly cited research papers within {\AIML} fields
shows that they contain many potentially harmful implicit biases and assumptions,
as well as an inherent selection and prioritisation of ethics aspects~\cite{Birhane_2022}.
The {accuracy} aspect
(indirectly represented as {performance})
is the most emphasised,
at the cost of considerably de-emphasising almost all other aspects.
The next most commonly prioritised quality is {generalisation},
which is haphazardly and inconsistently used as a proxy
for the {robustness} aspect in {\AIML} literature~\cite{Spratling_2024}.

The selection, prioritisation and trade-off resolution of AI ethics aspects
can occur at various points in the {\AIML} system development pipeline.
Without organisational policies and formal governance,
these can occur on an ad-hoc basis at the design and implementation levels,
and as such can be significantly affected by individual team members,
their knowledge and interpretation of Responsible AI issues,
personal preferences and bias \cite{Jakesch_2022,Varanasi_2023},
and lack of understanding of the effect of trade-offs on others~\cite{Orr_2020}.
On the other hand, explicit organisational policies may be under-developed
{\andor}~can lead to overly rigid adherence due to lack of flexibility~\cite{Rakova_2021,Varanasi_2023}.

We can consider a hierarchy of needs at three levels for addressing ethics trade-offs in Responsible AI systems:

\vspace{0.2ex}

{
\fontsize{9.9}{10.9}\selectfont
\begin{enumerate}[{$\bullet$},leftmargin=*]
\itemsep=0.4ex
\item
\textbf{Societal level}.
Governments, industry bodies and regulators determine high-level principles,
standards and regulations for {\AIML} system development and use~\cite{DISER_2020,UN_AI_2023}.
These are influenced by cultural norms, values and existing legislation. 

\item
\textbf{Organisational level}.
%Acknowledge that a set of trade-offs exists, rather than approaching them ad-hoc~\cite{Sanderson_2023b};
Instead of approaching trade-offs ad-hoc, acknowledge that a set of trade-offs exists~\cite{Sanderson_2023b};
ensure~designers and developers are aware of these trade-offs;
create frameworks and procedures for dealing with trade-offs
that are aligned with organisational, societal and regulatory expectations \cite{Rakova_2021}.

\item
\textbf{Practitioner level}. 
Prevent personal bias going into \mbox{trade-off} decisions;
be aware that there is more than one point of view~\cite{Jakesch_2022}
and that there may be implications for various groups~\cite{Orr_2020}.
Developers of {\AIML} systems action accepted frameworks
like risk assessments, justifications and patterns~\cite{Lu_2024},
in order to translate principles into practice.

\end{enumerate}
}

\begin{table*}[!tb]
\centering
\caption
  {%
  \rm\normalsize
  Summary of the high-level AI ethics principles proposed by the Australian Government~\cite{DISER_2020}.
  }
\label{tab:ai_ethics_principles}
\includegraphics[width=0.94\textwidth]{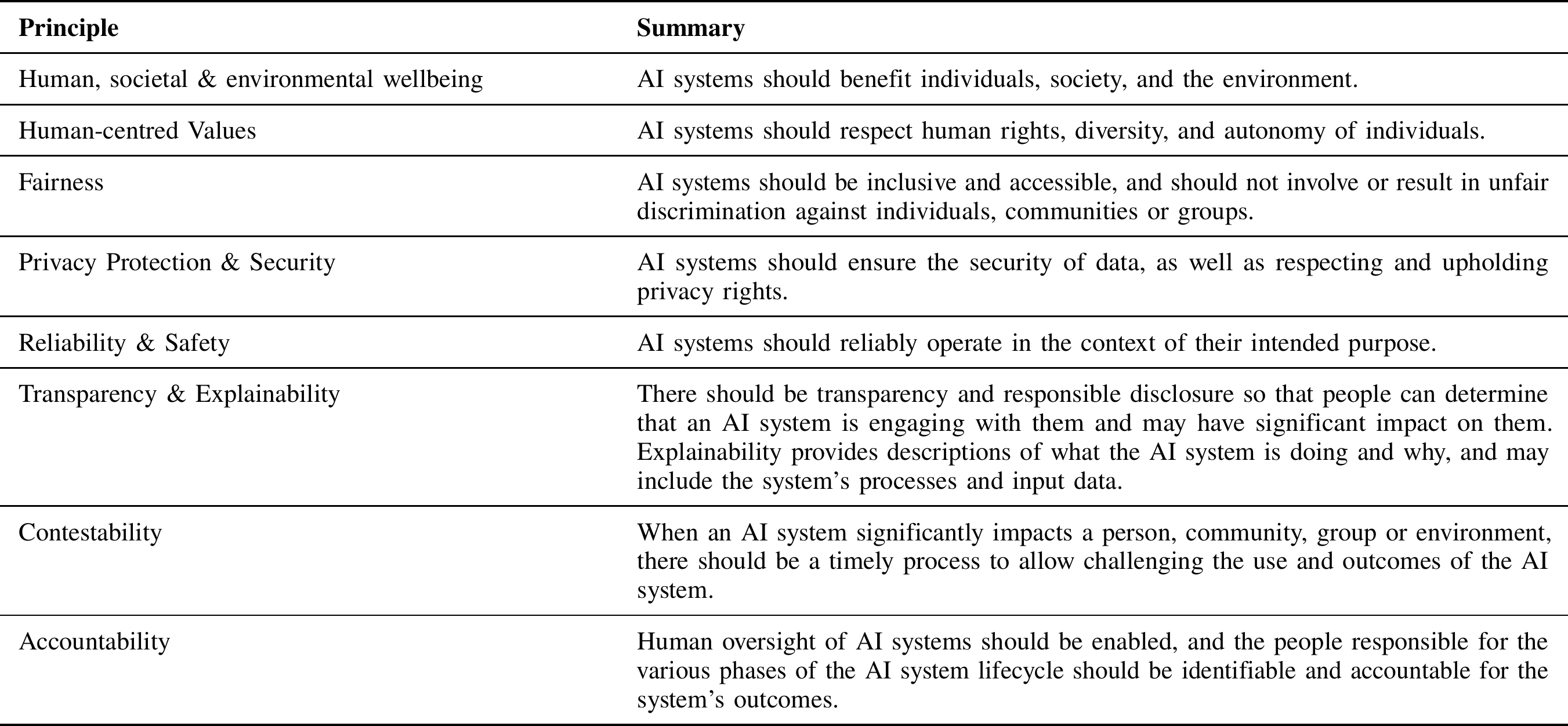}\\
%\vspace{1ex}
\end{table*}

\noindent
Given the above considerations,
an important next step 
is hence the development of frameworks {\andor} guidelines
to provide approaches to manage tensions and resolve trade-offs between AI ethics aspects,
in order to facilitate the design and implementation of well-rounded Responsible AI systems.

To that end, we summarise and analyse various approaches to addressing trade-offs in Sec.~\ref{sec:approaches},
noting their advantages and disadvantages.
We discuss the overall properties of the examined approaches in Sec.~\ref{sec:discussion},
and propose a multi-step framework that draws on the gained insights
as well as the needs at societal, organisational and practitioner levels.
Concluding remarks are given in Sec.~\ref{sec:conclusion}.

\begin{table}[!b]
\vspace{-3.5ex}
\hrule
\vspace{1ex}
\def\mydiagcell{\diagbox[linewidth=0.1pt]{~}{~}}
%\centering
\caption
  {
  Matrix of observed interactions between AI~ethics aspects, summarised from \cite{Sanderson_2023b}.
  \CIRCLE~=~trade-off;
  \Circle~=~synergistic~interaction;
  \LEFTcircle~=~mixed or context dependent interaction.
  }
\label{tab:ethics_interactions}
\vspace{-3.5ex}
\begin{tabular}{cc|c|c|c|c|c|c}
{~}
  & \rot{accuracy} & \rot{robustness} & \rot{fairness} & \rot{privacy}  & \rot{explainability} & \rot{transparency}
\\
\cline{2-7}
\multicolumn{1}{c|}{accuracy}
  & \mydiagcell    & \LEFTcircle      & \CIRCLE        & \CIRCLE        & \CIRCLE              & ~           \\
\cline{2-7}
\multicolumn{1}{c|}{robustness}
  & \LEFTcircle    & \mydiagcell      & \LEFTcircle    & \Circle        & ~                    & \LEFTcircle \\
\cline{2-7}
\multicolumn{1}{c|}{fairness}
  & \CIRCLE        & \LEFTcircle      & \mydiagcell    & \CIRCLE        &                      & \LEFTcircle \\
\cline{2-7}
\multicolumn{1}{c|}{privacy}
  & \CIRCLE        & \Circle          & \CIRCLE        & \mydiagcell    &                      & \LEFTcircle \\
\cline{2-7}
\multicolumn{1}{c|}{explainability}
  & \CIRCLE        &                  &                &                & \mydiagcell         & \Circle      \\
\cline{2-7}
\multicolumn{1}{c|}{transparency}
  &                & \LEFTcircle      & \LEFTcircle    & \LEFTcircle    & \Circle             & \mydiagcell  \\
\cline{2-7}
\end{tabular}
%\vspace{-3ex}
\end{table}

\section{Approaches for Resolving Trade-offs}
\label{sec:approaches}
\vspace{-1ex}

\subsection{Dominant Aspects}
\label{sec:approach_dom_asp}
\vspace{-0.5ex}

A blunt and straightforward approach to resolve tensions between ethics aspects
is to select the most dominant or pertinent aspect in a given context.
The prioritisation of aspects (eg.~accuracy over privacy)
can be driven by how difficult or costly it is to implement a given aspect within an {\AIML} system,
{\andor} internal organisational needs (eg.~regulatory compliance),
{\andor} the preferences of end users of the {\AIML} systems \cite{Jakesch_2022}.

The advantage of this approach is its overall simplicity and the low degree of required effort.
A~major disadvantage is that this is a winner-takes-all approach,
which leaves no room to devise balanced trade-offs and take into account nuance,
which in turn implies that a thorough evaluation of associated risks and benefits is not performed.
This approach is hence consistent with the pejorative notions of \textit{ethical lip service}~\cite{McLennan_2020}
and \textit{ethics washing}~\cite{Bietti_2020},
where only minimal effort is expended to address ethical issues that emerge in {\AIML} systems.

\subsection{Risk Reduction via Aspect Infringement and Amelioration}
\label{sec:approach_risk_red}

An indirect approach to resolving trade-offs is via prioritisation of ethics aspects, as proposed in \cite{Bruschi_2023}.
This involves a multi-step strategy to reduce the operational risk of {\AIML} systems,
summarised as follows.
First, an undesirable operational event in a given {\AIML} system is identified (eg.~a specific failure),
through a risk assessment matrix
that takes into account the likelihood of the event and associated degree of loss.
Secondly, the risk assessment matrix is expanded to allow degrees of infringement (de-prioritisation)
of given ethics aspects in order to reduce the risk of undesirable events.
Lastly, additional safeguards are put into place with the intention to ameliorate 
the infringement of the affected ethics aspects.

As per the example given in \cite{Bruschi_2023}, 
an AI-based user authentication system 
(eg.~to prevent unauthorised access to bank accounts)
can be made more accurate {\andor} more robust
(eg.~less susceptible to impersonation attacks)
by requiring the use of more personally identifiable information (eg.~face images~\cite{Cardinaux_2006}).
However, the use of such information ``infringes'' the privacy aspect.
The infringement is then ameliorated through further security measures to protect the information
and to comply with applicable laws such as EU's General Data Protection Regulation (GDPR) \cite{GDPR_2016_679}.

An advantage of the above multi-step strategy is that the prioritisation of ethics aspects
is driven by system-specific requirements and takes into account the context of system operation.
The disadvantage is that the risk assessment step may be error-prone
(eg. missing undesirable events, unreliable estimation of likelihood and loss),
and may require expert knowledge
which is beyond the level available to the design and development teams.
Furthermore, it may not be possible to adequately ameliorate the infringement of applicable ethics aspects.
Lastly, the incorporation of ethics aspects is done during later stages of {\AIML} system design,
which can be interpreted as treating the aspects as add-ons,
rather than taking them into account from the very outset.

\subsection{Trade-Off Analysis in Requirements Engineering}
\label{sec:approach_req_eng}

\begin{figure}[!b]
\centering
\includegraphics[width=1\columnwidth]{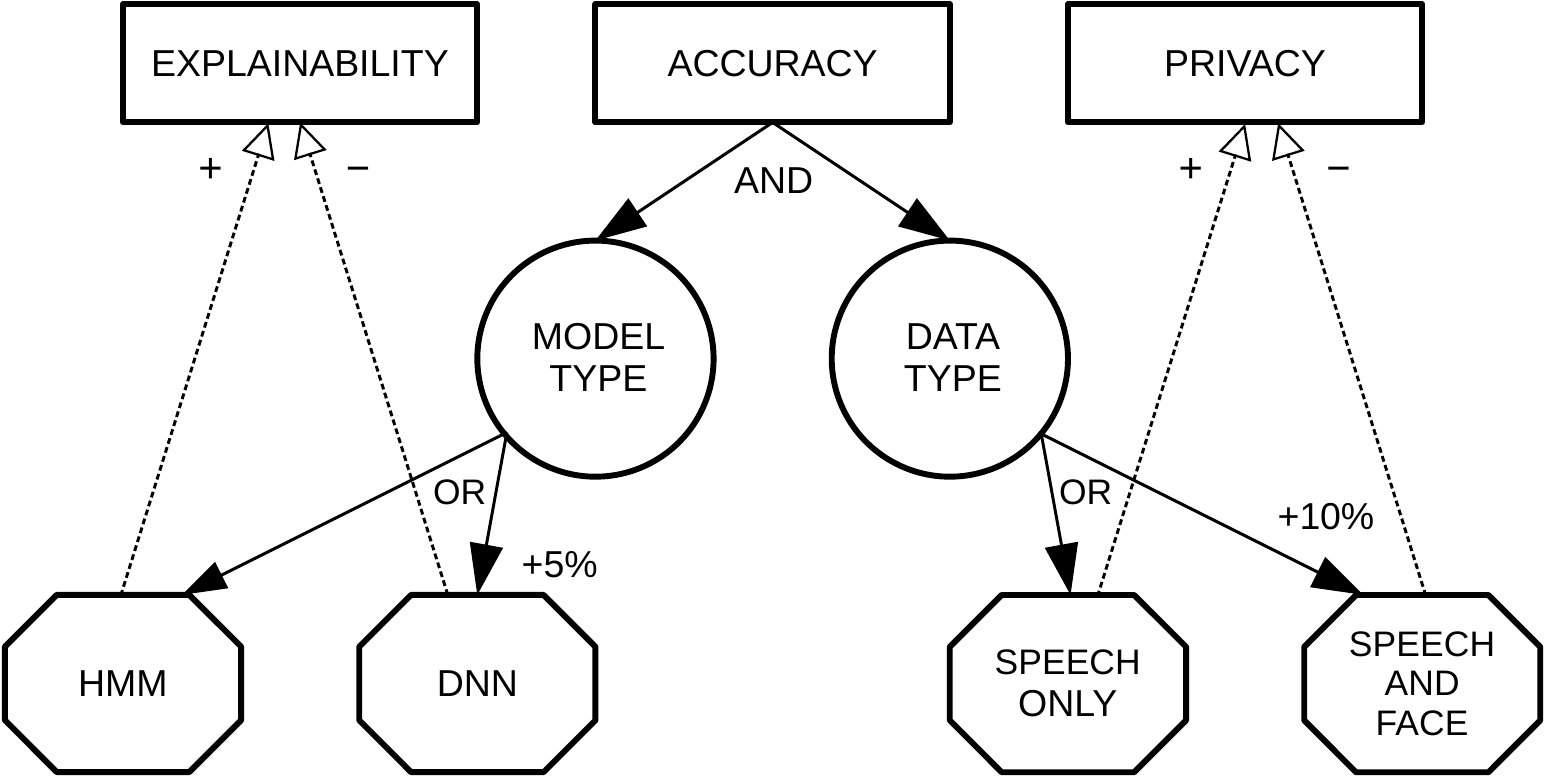}
\caption
  {
  An example of graphical trade-off analysis, adapted from~\cite{Maalej_2023}.
  For an AI based authentication system using biometrics,
  the consideration involves
  two ML models (HMM and DNN)
  and two data types (speech only, and speech in conjunction with face).
  }
\label{fig:tradeoff_analysis}
\end{figure}

Trade-off analysis techniques used in \textit{requirements engineering} 
\cite{Ahmad_2023,Horkoff_2019,Lu_CAIN_2022}
may be applicable for determining how each ethics aspect affects an AI system under construction
and for addressing the interplay between ethics aspects \cite{Koshiyama_2022,Maalej_2023}.
Each applicable ethics aspect is listed,
followed by listing possible ML models and data types that may be suitable for implementing an AI system.
Linkages between the ethics aspects and system components (model types and data types) are then graphically noted,
in conjunction with their positive or negative effect
(either in a quantitative or qualitative manner).

An illustrative example of the above trade-off analysis approach is shown in Fig.~\ref{fig:tradeoff_analysis}.
The AI system under construction is a biometric user authentication system employing speech {\andor} face data,
with the overall goal to increase security of banking services.
Accuracy of the AI system is driven by two main components: model type and data type~\cite{Bishop_2006}.
Two ML models are considered: 
Hidden Markov Model (HMM)~\cite{Cardinaux_2006} 
and Deep Neural Network (DNN)~\cite{Bai_2021}.
Furthermore, two data types are considered: speech only, and speech in conjunction with face data.
In this example, using DNN can increase the accuracy by 5\%,
but at the cost of reduced explainability in comparison to HMM.
Using speech and face data over using speech data alone can increase accuracy by 10\%,
though this negatively affects the privacy aspect as more personally identifiable information is used.

This type of trade-off analysis techniques can be considered as part of the design phase,
where many possible implementations of an overall {\AIML} system are explored.
The techniques can also be considered as part of the documentation phase,
where the trade-offs are explicitly documented (rather than left as tacit knowledge),
along with discussions on the practical pros and cons of each possible implementation.
This ties in with the dimension of justification suggested in \cite{Bleher_2023},
where the justification provides a context-specific rationale
for the drivers behind giving more weight to one aspect over another.

An advantage of this approach is that many possible trade-offs
can be explicitly shown and considered at the same time,
and the consideration is done as part of the design phase.
A~disadvantage is that the graphical representation can become quite complex
when more model types, data types and ethics aspects are considered.
Furthermore, multiple graphs may be required,
depending on the complexity of the ML pipeline~\cite{Xin_2021}.

\subsection{Quantitative Ranking of Trade-Off Solutions}
\label{sec:approach_quant_rank}

Inspired by~\cite{Loi_2021}, a ranking approach can be used to choose trade-off solutions.
Given several technical solutions to a given trade-off,
such as a set of possible ML models with various degrees of explainability
for the accuracy/explainability trade-off~\cite{Petkovic_2023},
each solution is ranked according to an overall score.

The overall score is a weighted convex linear combination of a set of normalised sub-scores~\cite{Strang_2023},
with each sub-score representing how well a given solution addresses a desired characteristic
pertinent to the trade-off at hand.
The set of characteristics can range from purely pragmatic (eg.~complexity of the ML model),
to various philosophical positions (such as utilitarianism and egalitarianism~\cite{Loi_2021}).
The weighting of the characteristics can be non-informative
(ie.~all weights are equal and sum to one),
or it can be based on the importance of each characteristic to an organisation
(ie.~weights are unequal and skewed towards focusing on a subset of characteristics).

The main advantage of this approach is that it aims to provide a quantitative procedure for resolving trade-offs,
and explicitly allows for the consideration of characteristics
that are important for the technical implementation of the {\AIML} system,
as well as wider organisational policies.
However, the flexibility can also be a disadvantage,
in that the selection of characteristics can be subject to errors,
and hence may require well-informed reasoning that may be beyond the capability of the practitioners {\andor} the organisation.
Furthermore, determining sub-scores and associated weights for the characteristics can be subjective,
especially when dealing with characteristics that are not easily quantifiable~\cite{Buijsman_2023}.
For example, it is non-trivial to represent the degree of egalitarianism as a precise numeric value.

\subsection{Specification and Balancing via Principlism}
\label{sec:approach_principlism}

Principlism is an influential approach in bioethics that uses a set of moral principles
to guide ethical decisions, such as respect for autonomy, nonmaleficence (avoid causing harm),
beneficence (promoting the welfare of others), and justice~\cite{Beauchamp_2013}.
The principlism approach has also been applied to cybersecurity ethics~\cite{Formosa_2021}.
The similarity with high-level AI ethics principles makes principlism a useful approach to draw on
when considering how to address trade-offs between the underlying AI ethics aspects, 
and for identifying the limitations of using a principlist approach to AI ethics~\cite{Mittelstadt_2019,Seger_2022}.

Principlism uses two approaches to bridge the gap between abstract principles and addressing individual 
cases~\cite{Beauchamp_2013}: (i)~specification, and (ii)~balancing.
Specification elaborates on the principles to describe how individual cases are relevant to a specific principle
or to the underlying ethical aspect behind it. For example, the principle of justice may be further specified 
by a rule that prohibits using ethnicity or gender as a basis for distributing access to resources~\cite{Shea_2020}.
AI ethics principles may also come with brief descriptions of how each principle may be applied~\cite{DISER_2020}.
However, such elaborations of basic principles will not cover all the possible cases,
and will not remove all the potential conflicts between them~\cite{Beauchamp_2013}.

When ethics principles or their underlying aspects give conflicting recommendations,
the principlism approach provides six conditions
that any balancing or trade-off must meet~\cite{Beauchamp_2013},
as summarised below:

%\begin{enumerate}[{1.},leftmargin=*]
\begin{enumerate}[{1.}]

\item
A stronger justification can be given for prioritising one aspect over another.

\item
The purpose of overriding a given aspect has a realistic chance of being achieved.

\item
There are no alternatives to this trade-off that are morally preferable.

\item
The overridden aspect is infringed to the smallest extent possible to achieve the purpose of overriding it.

\item
The negative effects of overriding the aspect are minimised.

\item
Those affected by this trade-off are treated impartially.

\end{enumerate}

\noindent
The justifications for prioritising a given aspect over another (condition~1) may be grounded in 
practical limitations or in ethical theory. Such justifications must still comply with the other five 
conditions.
Ethical concepts or theories that may ground such trade-offs include:

\begin{itemize}

\item
Proportionality: the methods used should be appropriate for the problem the {\AIML} system is intended to solve,
and should have a minimal impact on the system's compliance with the other aspects~\cite{Karliuk_2023}.

\item
Benefit to the least-advantaged (maximin):
decisions should be made based on maximising the benefits to the worst-off~\cite{Rawls_1971,Woodgate_2024}.

\item
Utilitarianism{\hspace{0.1ex}}/{\hspace{0.1ex}}Consequentialism:
decisions should be made based on maximising the utility or total benefits to all those affected by them~\cite{Woodgate_2024}.

\end{itemize}

\noindent
There are several advantages to adopting a principlist approach to addressing AI ethics concerns.
A set of ethics principles offers developers a common vocabulary for discussing ethics issues~\cite{Seger_2022}.
Each principle serves as a starting point for further ethical reflection and discussion
about how it may apply to a specific AI application.
The explicit statement of principles can also encourage cultural shifts within professions
towards prioritising the values and goals that they express~\cite{Seger_2022}. 
They may serve to establish ethical norms and change the values
that are prioritised within professional communities~\cite{Seger_2022}.

However, the principlism approach also has several disadvantages.
While the six conditions described above provide guidelines for how to make trade-offs between ethics aspects,
how developers respond to these conditions is left up to the developers' judgement~\cite{Flynn_2022}.
One set of developers may take a consequentialist approach to justifying the trade-offs they make,
while another may favour using maximin.
As a result, various developers may make different trade-offs when faced with the same conflict between ethics aspects. 

Furthermore, {\AIML} system development occurs in a significantly different context to medicine,
where principlism has been influential as an approach to ethics~\cite{Mittelstadt_2019}.
Unlike medicine, AI development does not have common goals, a~long professional history that has 
clear descriptions of the ethical duties expected of practitioners, a~relative lack of methods for 
interpreting principles into practice, and the relative lack of professional accountability~\cite{Mittelstadt_2019}.
These differences may limit the effectiveness of ethics principles for influencing AI design and governance~\cite{Mittelstadt_2019}.

\vspace{-1ex}
\section{Discussion and Recommendations}
\label{sec:discussion}

In the preceding section we examined five approaches 
that can be incorporated into the {\AIML} system design process
with the aim of addressing tensions between common AI ethics aspects.
From across the covered approaches we can extract five properties:
(1)~prioritisation of ethics aspects based on risk assessment/context specific assessment;
(2)~proactive analysis at start of design process;
(3)~de-prioritisation of an ethics aspect in favour of another;
(4)~analysis requires deliberation {\andor} resources to enact;
(5)~additional considerations required to ameliorate side-effects.
Table~\ref{tab:properties} indicates the presence and absence of the above properties
for each of the approaches.

\begin{table}[!b]
\vspace{-1ex}
\centering
\hrule
\vspace{1ex}
%\rule{1\columnwidth}{0.5pt}%
\renewcommand{\arraystretch}{1.2}
\small
\caption
  {
  Properties of approaches in Sec.~\ref{sec:approaches} for addressing trade-offs.
  (1)~=~prioritisation of ethics aspects based on risk assessment/context specific assessment;
  (2)~=~proactive analysis at start of design process;
  (3)~=~de-prioritisation of an ethics aspect in favour of another;
  (4)~=~requires deliberation/resources to enact;
  (5)~=~additional considerations required to ameliorate side-effects.
  \CIRCLE~=~property present;
  \Circle~=~property absent.
  }
\label{tab:properties}
\vspace{-1ex}
\begin{tabular}{rl|cccccc}
\hline
\multirow{2}{*}{\textbf{Approach}} & ~ & \multicolumn{5}{c}{\textbf{Property}} \\ 
~ & ~ & (1) & (2) & (3) & (4) & (5) \\
\hline
Dominant Aspects     & \hspace{-2ex}(Sec.~\ref{sec:approach_dom_asp})     & \Circle & \Circle & \CIRCLE & \Circle & \Circle \\
Risk Reduction       & \hspace{-2ex}(Sec.~\ref{sec:approach_risk_red})    & \CIRCLE & \Circle & \CIRCLE & \CIRCLE & \CIRCLE \\
Requirements Eng.    & \hspace{-2ex}(Sec.~\ref{sec:approach_req_eng})     & \CIRCLE & \CIRCLE & \CIRCLE & \CIRCLE & \Circle \\
Quantitative Ranking & \hspace{-2ex}(Sec.~\ref{sec:approach_quant_rank})  & \CIRCLE & \Circle & \CIRCLE & \CIRCLE & \Circle \\
Principlism          & \hspace{-2ex}(Sec.~\ref{sec:approach_principlism}) & \CIRCLE & \Circle & \CIRCLE & \CIRCLE & \CIRCLE \\
\hline
\end{tabular}
\end{table}

There is no single approach that is likely to be appropriate for all organisations, {\AIML} systems, or applications. 
In order to address this shortcoming,
we propose a layered framework to introduce into the {\AIML} system design pipeline,
which is enacted at a practitioner level while being supported by organisational and societal influence.

Guidance of the design process and high-level decision making at a \textit{societal level}
can be informed by scientific literature, AI ethics principles, standards and regulations.
These dimensions, however, do not take into account context,
which is a key consideration in design decisions when managing tensions.
Context includes the purpose of the {\AIML} system, the groups of people impacted by the system,
and the level of understanding about the outputs of the system that are required~\cite{Arrieta_2020}.
At an \textit{organisational level},
policies and governance can be more specific to the general context of an AI application~\cite{Blanchard_2024}, 
but will generally be agnostic to the nuanced details required to make design decisions on a case-by-case~basis.
At the \textit{practitioner~level}, where the resolution of tensions is specific to the context of the AI model,
risk assessments as well as design tools and justifications can be used. 

We can use insights from the approaches described in Sec.~\ref{sec:approaches}
to build a multi-step framework around managing trade-offs and tensions between ethics aspects.
The proposed framework is comprised of three main components:
(i)~proactive identification,
(ii)~prioritisation and weighting,
(iii)~justification and documentation.
The components are elucidated below.

\textbf{Proactive identification}.
A proactive and structured assessment at the beginning of the {\AIML} design pipeline
provides a framework to examine and define the context and purpose of the {\AIML} system.
It also encourages a proactive approach for the identification of potential tensions between ethics aspects 
as well as the consideration of the methodology required to resolve them.
The \textit{requirements engineering} approach (Sec.~\ref{sec:approach_req_eng})
encompasses proactive design consideration where ethics aspects,
models and data types are examined, and the associated tensions and solutions are~explored~\cite{Maalej_2023}.
Similarly, a Value Sensitive Design~(VSD) approach exemplifies a proactive approach
to applying high-level ethics principles in practice~\cite{Bleher_2023}.
VSD is an iterative and multidisciplinary process where ethical considerations
are incorporated into the design process from the start \cite{Boyd_2022}.
Other approaches consider tensions between ethics aspects as they arise in the design process.
For example, ranking of trade-off solutions (Sec.~\ref{sec:approach_quant_rank}) would routinely occur
when tensions arise and can be used in addition to a prospective risk analysis.
Specifying what the ethics principles mean in the context of a specific {\AIML} project
(such as in the \textit{principlism} approach in Sec.~\ref{sec:approach_principlism})
may also identify possible tensions between ethics aspects early in the design process
and hence allow them to be addressed.

\textbf{Prioritisation and weighting}.
The nature of resolving tensions between ethics aspects that arise when developing {\AIML} systems
is that trade-offs must be made,
resulting in 
(a)~one or more ethics aspects being prioritised over others,
{\andor}
(b)~a weighted (balanced) combination of selected ethics aspects.
The least critical technique to apply is the \textit{dominant aspects} approach (Sec.~\ref{sec:approach_dom_asp}),
where prioritisation is made based on one-dimensional decisions such as difficulty or cost.
In contrast, the other approaches listed in Sec.~\ref{sec:approaches}
require prioritisation based on a degree of context-based assessment.
To this end, the first step in prioritising one ethics aspect over another is a consideration of the broader context.
This can be done in a qualitative manner to selectively
introduce trade-offs in order to reduce identified operational risks,
as per the \textit{risk reduction} approach (Sec.~\ref{sec:approach_risk_red}).
Alternatively, the \textit{quantitative ranking} approach (Sec.~\ref{sec:approach_quant_rank}) can be used,
where various solutions are ranked according to desired characteristics.
Such rankings can also inform the responses to the six conditions
specified for resolving conflicts between ethics aspects
used by the \textit{principlism} approach (Sec.~\ref{sec:approach_principlism}).
A hybrid approach is used by the \textit{requirements engineering} approach (Sec.~\ref{sec:approach_req_eng}),
where both quantitative and qualitative measurements are employed.

\textbf{Justification and documentation}.
Explicability and transparency are critical aspects of responsible design \cite{Arrieta_2020}.
As part of that, documentation and reasoning are essential 
concerning the design decisions where trade-offs have been made.
Justification provides a context-specific rationale for the drivers
behind giving more weight to one ethics aspect over another~\cite{Bleher_2023}.
The \textit{requirements engineering} approach (Sec.~\ref{sec:approach_req_eng})
can provide justification focused on practical aspects of {\AIML} system functionality,
while the \textit{quantitative ranking} approach (Sec.~\ref{sec:approach_quant_rank}) 
can take into account dimensions that go beyond immediate practical aspects.
The \textit{principlism} approach (Sec.~\ref{sec:approach_principlism})
takes a further step and provides a strong framework to develop justifications,
where prescriptive conditions are explored in the process
and can then be used to document and justify trade-off decisions.
For example, documentation that addresses how the six conditions
described in Sec.~\ref{sec:approach_principlism}
are being met through considered design decisions,
so that there is accountability for how the design decisions were made.
These explanations (and the impacts of the resulting trade-offs)
may also inform future decisions about resolving ethical tensions that arise in other projects.

\vspace{-2ex}
\section{Concluding Remarks}
\label{sec:conclusion}

While progress has been made in the application of high-level AI ethics principles 
in the design and implementation of {\AIML} systems~\cite{Morley_2023,Sanderson_2023a,Stix_2021},
there are still notable areas of concern, 
including a theory-practice gap in how to manage tensions
that arise between commonly accepted AI ethics aspects that underpin the principles~\cite{Bleher_2023,Sanderson_2023b}.

In this work, we have covered five approaches for addressing the tensions via trade-offs,
ranging from rudimentary to quite complex.
The approaches mainly differ in the types of considered context,
the scope of each context,
the associated methods for qualitatively {\andor} quantitatively measuring each context,
and how each trade-off decision is justified.
None of these approaches is likely to be appropriate for all organisations, {\AIML} systems, or applications.

In response, we have proposed a framework for {\AIML} system developers for use in the design pipeline
that draws on the various strengths of the covered approaches.
The framework has three main components:
(i)~proactive identification,
(ii)~prioritisation and weighting,
(iii)~justification and documentation.
At the start of the {\AIML} design pipeline, 
potential tensions between ethics aspects are identified,
and consideration is given to the methodology for resolving them.
The tensions between the identified aspects are then addressed
through prioritisation and weighting,
using one or more of the covered approaches;
both practical and organisational requirements can be taken into account.
Each trade-off decision is justified and documented,
aiding transparency and accountability,
as well as adding to the pool of organisational knowledge.

{\AIML} systems built with rudimentary {\andor} shallow ethical assessments 
are unlikely to be robust against potential legal and regulatory challenges.
Employing a proactive and dynamic assessment method across the full {\AIML} system pipeline
(including the context of the system's application),
such as the framework proposed here,
is more likely to yield well-rounded systems
that are appropriately designed and implemented for their regulatory environment.

\def~{\,}  % refine ~ character to be a shorter non-breaking space; https://texfaq.org/FAQ-activechars

\bibliographystyle{ieee_mod}
\balance
\bibliography{references}

\begin{thebibliography}{10}\interlinepenalty=10000\itemsep=0.5ex

\bibitem{Ahmad_2023}
K.~Ahmad, M.~Abdelrazek, C.~Arora, et~al.
\newblock Requirements engineering framework for human-centered artificial
  intelligence software systems.
\newblock {\em Applied Soft Computing}, 143:110455, 2023.

\bibitem{DISER_2020}
{Australian Government (Department of Industry, Science and Resources)}.
\newblock {Australia's Artificial Intelligence Ethics Framework}, 2019.
\newblock Accessed 28~Nov~2023. {URL:}
  {https://industry.gov.au/data-and-publications/australias-artificial-intelligence-ethics-framework/}.

\bibitem{Bai_2021}
Z.~Bai and X.-L. Zhang.
\newblock Speaker recognition based on deep learning: An overview.
\newblock {\em Neural Networks}, 140:65--99, 2021.

\bibitem{Arrieta_2020}
A.~Barredo~Arrieta et~al.
\newblock Explainable artificial intelligence ({XAI}): Concepts, taxonomies,
  opportunities and challenges toward responsible~{AI}.
\newblock {\em Information Fusion}, 58:82--115, 2020.

\bibitem{Beauchamp_2013}
T.~L. Beauchamp and J.~F. Childress.
\newblock {\em Principles of Biomedical Ethics, Seventh Edition}.
\newblock Oxford University Press, 2013.

\bibitem{Bietti_2020}
E.~Bietti.
\newblock From ethics washing to ethics bashing: a view on tech ethics from
  within moral philosophy.
\newblock In {\em ACM Conference on Fairness, Accountability, and Transparency
  (FAccT)}, pages 210--219, 2020.

\bibitem{Birhane_2022}
A.~Birhane, P.~Kalluri, D.~Card, W.~Agnew, R.~Dotan, and M.~Bao.
\newblock The~values encoded in machine learning research.
\newblock In {\em ACM Conf.~Fairness, Accountability, and Transparency
  (FAccT)}, pages 173--184, 2022.

\bibitem{Bishop_2006}
C.~M. Bishop.
\newblock {\em Pattern Recognition and Machine Learning}.
\newblock Springer, 2006.

\bibitem{Blanchard_2024}
A.~Blanchard, C.~Thomas, and M.~Taddeo.
\newblock Ethical governance of artificial intelligence for defence: normative
  tradeoffs for principle to practice guidance.
\newblock {\em AI \& Society}, 2024.

\bibitem{Bleher_2023}
H.~Bleher and M.~Braun.
\newblock Reflections on putting {AI} ethics into practice: How three {AI}
  ethics approaches conceptualize theory and practice.
\newblock {\em Science and Engineering Ethics}, 29(3), 2023.

\bibitem{Boyd_2022}
K.~Boyd.
\newblock Designing up with value-sensitive design: Building a field guide for
  ethical {ML} development.
\newblock In {\em ACM Conference on Fairness, Accountability, and
  Transparency}, pages 2069--2082, 2022.

\bibitem{Bruschi_2023}
D.~Bruschi and N.~Diomede.
\newblock A framework for assessing {AI} ethics with applications to
  cybersecurity.
\newblock {\em AI Ethics}, 3(1):65--72, 2023.

\bibitem{Buijsman_2023}
S.~Buijsman.
\newblock Navigating fairness measures and trade-offs.
\newblock {\em AI and Ethics}, 2023.

\bibitem{Cardinaux_2006}
F.~Cardinaux, C.~Sanderson, and S.~Bengio.
\newblock User authentication via adapted statistical models of face images.
\newblock {\em IEEE Transactions on Signal Processing}, 54(1):361--373, 2006.

\bibitem{GDPR_2016_679}
{European Union}.
\newblock {Regulation (EU) 2016/679 ... (General Data Protection Regulation)}.
\newblock Official Journal of the European Union L 119, 4~May~2016.

\bibitem{Fjeld_2020}
J.~Fjeld, N.~Achten, H.~Hilligoss, A.~C. Nagy, and M.~Srikumar.
\newblock Principled artificial intelligence: Mapping consensus in ethical and
  rights-based approaches to principles for {AI}.
\newblock {Berkman Klein Center for Internet \& Society at Harvard University},
  \textit{Research Publication No.~2020-1}, 2020.

\bibitem{Flynn_2022}
J.~Flynn.
\newblock {Theory and Bioethics}.
\newblock In {\em The {Stanford} Encyclopedia of Philosophy}. Metaphysics
  Research Lab, Stanford University, 2022.

\bibitem{Formosa_2021}
P.~Formosa, M.~Wilson, and D.~Richards.
\newblock A principlist framework for cybersecurity ethics.
\newblock {\em Computers \& Security}, 109:102382, 2021.

\bibitem{Holm_2019}
E.~Holm.
\newblock In defense of the black box.
\newblock {\em Science}, 364(6435):26–27, 2019.

\bibitem{Horkoff_2019}
J.~Horkoff.
\newblock Non-functional requirements for machine learning: Challenges and new
  directions.
\newblock In {\em IEEE International Requirements Engineering Conference},
  2019.

\bibitem{Huang_2023}
C.~Huang, Z.~Zhang, B.~Mao, and X.~Yao.
\newblock An overview of artificial intelligence ethics.
\newblock {\em IEEE Transactions on Artificial Intelligence}, 4(4):799--819,
  2023.

\bibitem{Jakesch_2022}
M.~Jakesch, Z.~Buçinca, S.~Amershi, and A.~Olteanu.
\newblock How different groups prioritize ethical values for responsible {AI}.
\newblock In {\em ACM Conf.~Fairness, Accountability, and Transparency
  (FAccT)}, pages 310--323, 2022.

\bibitem{jobin2019global}
A.~Jobin, M.~Ienca, and E.~Vayena.
\newblock The global landscape of {AI} ethics guidelines.
\newblock {\em Nature Machine Intelligence}, 1(9):389--399, 2019.

\bibitem{Karliuk_2023}
M.~Karliuk.
\newblock Proportionality principles for the ethics of artificial intelligence.
\newblock {\em AI and Ethics}, 3(3):985--990, 2023.

\bibitem{Koshiyama_2022}
A.~Koshiyama, E.~Kazim, and P.~Treleaven.
\newblock Algorithm auditing: Managing the legal, ethical, and technological
  risks of artificial intelligence, machine learning, and associated
  algorithms.
\newblock {\em Computer}, 55(4):40--50, 2022.

\bibitem{Kozodoi_2022}
N.~Kozodoi, J.~Jacob, and S.~Lessmann.
\newblock Fairness in credit scoring: Assessment, implementation and profit
  implications.
\newblock {\em European Journal of Operational Research}, 297(3):1083--1094,
  2022.

\bibitem{Loi_2021}
M.~Loi and M.~Christen.
\newblock Choosing how to discriminate: navigating ethical trade-offs in fair
  algorithmic design for the insurance sector.
\newblock {\em Philosophy \& Technology}, 34(4):967--992, 2021.

\bibitem{Lu_CAIN_2022}
Q.~Lu, L.~Zhu, X.~Xu, J.~Whittle, and Z.~Xing.
\newblock Towards a roadmap on software engineering for responsible {AI}.
\newblock In {\em International Conference on {AI} Engineering: Software
  Engineering for {AI}}, pages 101--112, 2022.

\bibitem{Lu_2024}
Q.~Lu, L.~Zhu, X.~Xu, J.~Whittle, D.~Zowghi, and A.~Jacquet.
\newblock \mbox{Responsible} {AI} pattern catalogue: A collection of best
  practices for {AI} governance and engineering.
\newblock {\em ACM Computing Surveys}, 56(7), 2024.

\bibitem{Maalej_2023}
W.~Maalej, Y.~D. Pham, and L.~Chazette.
\newblock Tailoring requirements engineering for responsible {AI}.
\newblock {\em Computer}, 56(4):18–27, 2023.

\bibitem{McLennan_2020}
S.~McLennan, A.~Fiske, L.~A. Celi, R.~Müller, J.~Harder, K.~Ritt, et~al.
\newblock An embedded ethics approach for {AI} development.
\newblock {\em Nature Machine Intelligence}, 2(9):488--490, 2020.

\bibitem{Mittelstadt_2019}
B.~Mittelstadt.
\newblock Principles alone cannot guarantee ethical {AI}.
\newblock {\em Nature Machine Intelligence}, 1(11):501--507, 2019.

\bibitem{Morley_2023}
J.~Morley, L.~Kinsey, A.~Elhalal, F.~Garcia, M.~Ziosi, and L.~Floridi.
\newblock Operationalising {AI} ethics: barriers, enablers and next steps.
\newblock {\em {AI} \& Society}, 38(1):411--423, 2023.

\bibitem{Orr_2020}
W.~Orr and J.~L. Davis.
\newblock Attributions of ethical responsibility by artificial intelligence
  practitioners.
\newblock {\em Information, Communication \& Society}, 23(5):719--735, 2020.

\bibitem{Peters_2020}
D.~Peters, K.~Vold, D.~Robinson, and R.~A. Calvo.
\newblock Responsible {AI} -- two frameworks for ethical design practice.
\newblock {\em IEEE Transactions on Technology and Society}, 1(1):34–47,
  2020.

\bibitem{Petkovic_2023}
D.~Petkovic.
\newblock It is not `accuracy vs. explainability' -- we need both for
  trustworthy {AI} systems.
\newblock {\em IEEE Transactions on Technology and Society}, 4(1):46--53, 2023.

\bibitem{Rakova_2021}
B.~Rakova, J.~Yang, H.~Cramer, and R.~Chowdhury.
\newblock Where {Responsible AI} meets reality.
\newblock {\em Proceedings of the ACM on Human-Computer Interaction},
  5(CSCW1):7:1--23, 2021.

\bibitem{Rawls_1971}
J.~Rawls.
\newblock {\em A Theory of Justice}.
\newblock Harvard University Press, 1971.

\bibitem{Sanderson_2023b}
C.~Sanderson, D.~Douglas, and Q.~Lu.
\newblock Implementing responsible {AI}: Tensions and trade-offs between ethics
  aspects.
\newblock In {\em International Joint Conference on Neural Networks}, 2023.

\bibitem{Sanderson_2023a}
C.~Sanderson, D.~Douglas, Q.~Lu, E.~Schleiger, J.~Whittle, J.~Lacey,
  G.~Newnham, S.~Hajkowicz, C.~Robinson, and D.~Hansen.
\newblock \scalebox{0.9665}{{AI} ethics principles in practice: perspectives of
  designers and developers}.
\newblock {\em IEEE Transactions on Technology and Society}, 4(2):171--187,
  2023.

\bibitem{Seger_2022}
E.~Seger.
\newblock In defence of principlism in {AI} ethics and governance.
\newblock {\em Philosophy \& Technology}, 35:45, 2022.

\bibitem{Shea_2020}
M.~Shea.
\newblock Principlism's balancing act: Why the principles of biomedical ethics
  need a theory of the good.
\newblock {\em Journal of Medicine and Philosophy}, 45:441--470, 2020.

\bibitem{Smit_2020}
K.~Smit, M.~Zoet, and J.~van Meerten.
\newblock A review of {AI} principles in practice.
\newblock In {\em Proc. Pacific Asia Conference on Information Systems}, 2020.

\bibitem{Spratling_2024}
M.~W. Spratling.
\newblock A comprehensive assessment benchmark for rigorously evaluating deep
  learning image classifiers.
\newblock \textit{arXiv:2308.04137~(v2)}, 2024.

\bibitem{Stix_2021}
C.~Stix.
\newblock Actionable principles for artificial intelligence policy: Three
  pathways.
\newblock {\em Science and Engineering Ethics}, 27(1):15, 2021.

\bibitem{Strang_2023}
G.~Strang.
\newblock {\em Introduction to Linear Algebra}.
\newblock Wellesley-Cambridge Press, 6th edition, 2023.

\bibitem{UN_AI_2023}
{United Nations}.
\newblock High-level advisory body on artificial intelligence, 2023.
\newblock Accessed 28~Nov~2023. {https://www.un.org/en/ai-advisory-body}.

\bibitem{Vakkuri_2020}
V.~Vakkuri, K.-K. Kemell, M.~Jantunen, and P.~Abrahamsson.
\newblock This is just a prototype: How ethics are ignored in software
  startup-like environments.
\newblock {\em Lecture Notes in Business Information Processing}, 383:195--210,
  2020.

\bibitem{Varanasi_2023}
R.~A. Varanasi and N.~Goyal.
\newblock It is currently hodgepodge: Examining {AI/ML} practitioners'
  challenges during co-production of responsible {AI} values.
\newblock In {\em ACM Conf. Human Factors in Computing Systems}, 2023.

\bibitem{Whittlestone_2019}
J.~Whittlestone, R.~Nyrup, A.~Alexandrova, and S.~Cave.
\newblock The role and limits of principles in {AI} ethics: Towards a focus on
  tensions.
\newblock In {\em AAAI/ACM Conference on AI, Ethics, and Society}, pages
  195--200, 2019.

\bibitem{Woodgate_2024}
J.~Woodgate and N.~Ajmeri.
\newblock Macro ethics principles for responsible {AI} systems: Taxonomy and
  directions.
\newblock {\em {ACM} Comput. Surv.}, 56(11), 2024.

\bibitem{Xin_2021}
D.~Xin, H.~Miao, A.~Parameswaran, and N.~Polyzotis.
\newblock Production machine learning pipelines: Empirical analysis and
  optimization opportunities.
\newblock In {\em ACM Int. Conf. Management of Data}, pages 2639--2652, 2021.

\end{thebibliography}

\end{document}